\documentclass[conference]{IEEEtran}
\usepackage{graphicx}
\usepackage{array}
\usepackage{color}
\usepackage{amsmath}
\usepackage{multirow}

\ifCLASSINFOpdf
\else
\fi
\usepackage{url}
\usepackage{units}
\usepackage{times}

\newcommand{\remove}[1]{}

\newtheorem{example}{Example}

\newcommand{\specialcell}[2][c]{%
	\begin{tabular}[#1]{@{}c@{}}#2\end{tabular}}

\begin{document}
	\title{Has the Online Discussion Been Manipulated? Quantifying Online Discussion Authenticity
		within Online Social Media \vspace{-0.3cm}}

	\author{\IEEEauthorblockN{Aviad Elyashar, Jorge Bendahan, and Rami Puzis}
		\IEEEauthorblockA{Telekom Innovation Laboratories \\
			Department of Software and Information Systems Engineering\\
			Ben-Gurion University of the Negev, Beer-Sheva, Israel\\
			Email: {\{aviade, jorgeaug\}}@post.bgu.ac.il, puzis@bgu.ac.il}

}
	
	\maketitle
	
	\begin{abstract}
		Online social media (OSM) has a enormous influence in today's world.
		Some individuals view OSM as fertile ground for abuse and use it to disseminate misinformation and political propaganda, slander competitors, and spread spam.
		The crowdturfing industry employs large numbers of bots and human workers to manipulate OSM and misrepresent public opinion.  
		The detection of online discussion topics manipulated by OSM \emph{abusers} is an emerging issue attracting significant attention.  
		In this paper we propose an approach for quantifying the authenticity of online discussions based on the similarity of OSM accounts participating in the discussion to known \emph{abusers} and \emph{legitimate accounts}. 
		Our method uses several similarity functions for the analysis and classification of OSM accounts. 
		The proposed methods are demonstrated using Twitter data collected for this study and previously published \emph{Arabic honeypot dataset}.  
		The former includes manually labeled accounts and \emph{abusers} who participated in crowdturfing platforms.  
		Evaluation of the topic's authenticity, derived from account similarity functions, shows that the suggested approach is effective for discriminating between topics that were strongly promoted by \emph{abusers} and topics that attracted authentic public interest. 
	\end{abstract}
	
	\IEEEpeerreviewmaketitle

	\section{Introduction}
	Online social media (OSM) allows people to share opinions and content, and in some cases, influence large segments of society~\cite{ratkiewicz2011truthy,lee2014dark,xiao2015detecting}. 
	Significant attention has been paid to trends emerging from OSM~\cite{asur2011trends}.
	Machiavellian individuals and organizations often attempt to harness the power of OSM in order to gain influence, damage the reputation of competitors, or spread political propaganda by financing disinformation campaigns~\cite{wang2012serf,lee2013crowdturfers,DBLP:conf/asunam/DickersonKS14}.
	Such campaigns are called to as \emph{crowdturfing}, stemming from crowdsourcing\footnote{Employing the Internet crowd to perform many micro-jobs.} and astroturfing\footnote{Hiding the sponsors of a message to make it appear genuine.}.
	
	In many cases crowdturfing campaigns utilize OSM accounts, operated by humans~\cite{wang2012serf,lee2013crowdturfers} or by computer programs known as bots~\cite{Chu:2010:TTH:1920261.1920265,ferrara2014rise}.
	These malicious accounts may cause serious damage. 
	For example, in 2014, tweets posted by socialbots influenced automated trading algorithms, causing a boost in the stock market prices of a tech company whose stock value jumped 200-fold, increasing the company's market value to five billion dollars in a matter of weeks~\cite{ferrara2014rise}. 
	In another case, Vietnamese officials employed a large number of crowdturfers to engage in online discussions to spread messages in support of the ruling political party's policies~\cite{lee2014dark}.
	More recently, fake news was reported to proliferate in the US and European political arena~\cite{kucharski2016post,khaldarova2016fake}.  
	All of these examples demonstrate the need for comprehensive solutions to tackle OSM manipulation by crowdturfers~\cite{simonite2011hidden}. 
	
	In this study, we strive to discriminate between authentic online discussions and crowdturfing campaigns.
	Currently, extensive research is being conducted in the area of detecting OSM abusers such as spammers, bots, trolls, etc~\cite{lee2013crowdturfers,DBLP:conf/asunam/DickersonKS14,ferrara2014rise}. 
	However, it is not always possible to identify an account as a human or bot, because of the prevalence of semi-automated accounts within OSM~\cite{chu2010tweeting}. 
	Similarly, it not always possible to differentiate between crowdturfers and grassroots efforts to promote an area of online discussion.  
	
	Further complicating matters, are casual crowdturfers -- OSM accounts that largely exhibit authentic behavior but occasionally participate in malicious manipulation of OSM discussions~\cite{lee2013crowdturfers,lee2015characterizing}.
	This type of behavior in which abusers attempt to appear as regular accounts as much as possible is known as camouflage~\cite{hooi2016fraudar,wu2017detecting}.
	Therefore, quantifying the level of authenticity of an OSM account remains challenging. 
	
	In this study, we propose several similarity functions for comparing OSM accounts participating in a particular online discussion to both confirmed OSM \emph{abusers} and known \emph{legitimate accounts}. 
	These similarity functions can be used to quantify the level of an account's authenticity, where \emph{abusers} and \emph{legitimate accounts} are placed on the opposite ends of the authenticity scale. 
	Accounts' authenticity can, in turn, be aggregated to quantify the authenticity of online discussions. 
	We demonstrate that the distribution of accounts' authenticity is different in topics that are prone to OSM manipulation and topics that attract authentic public interest.  	
	In order to choose the best similarity function we used three Twitter\footnote{https://twitter.com/} datasets, two of which were collected during this study and one which was made available by Morstatter et al.~\cite{morstatter2016new}.
	
	The contributions of this paper are:
	\begin{itemize}
		\item a collection of two Twitter datasets that include 605 manually labeled accounts, 1,006 abusers who participated in crowdturfing platforms, and 88,506 unlabeled accounts (the data is available upon request);
		\item an application of similarity functions for account type classification;  
		\item a visual representation of the account authenticity distribution within OSM discussions;
	\end{itemize}	
	The Credibility of online information has been discussed in~\cite{gayo2013predicting}, however as the researchers mention, their approach only considers shallow characteristics (e.g., number of followers). 
	They do not distinguish between accounts that have acquired a positive reputation in the past and those who have been spreading misinformation, spam, etc.
	On the other hand, our approach does exactly the opposite. 
	It analyzes all the content published by the OSM accounts in order to improve prediction of OSM account authenticity. 
	Intuitive visualization of the account authenticity distribution will help OSM analysts in identifying fake news, assessing public opinion about commercial products and services, identifying OSM manipulation in political campaigns, etc. 
	
	The rest of this paper is organized as follows: 
	in Section~\ref{background} we review well-known methods for the detection of \emph{abusers} and the concept of topic modeling.  
	In Section~\ref{sec:genuineness_of_social_discussion}, we explain our approach for quantifying the authenticity of online discussions using similarity functions, including the account labeling process (see Section~\ref{sec:account_labeling}), a full description of the similarity functions (see Section~\ref{sec:author_similarity_functions}), and the approach for measuring the authenticity of accounts and topics (see Section~\ref{sec:authenticity_of_accounts_and_topics}).     
	The evaluation of our method is addressed in Section~\ref{sec:evaluation}.
	More specially, the datasets we evaluated our method on are presented in Section~\ref{sec:datasets}, and the results of the evaluation carried out on the datasets are presented in Section~\ref{sec:account_classification}
	Section~\ref{ethical} includes ethical considerations, and we conclude the paper in Section~\ref{conclusions} with a summary and introduction to future work.
	
	\section{Related Work} 
	\label{background}
	
	In this section, we outline studies involving the identification of \emph{abusers} in OSM.
	In 2012, Cao et al.~\cite{cao2012aiding} proposed a method that is used to cluster accounts based on the similarity of posted URLs; once this is done, each cluster is classified as either malicious or not by extracting 
	behavioral and content features.

	In 2013, a method for the identification of crowdturfers on Twitter was presented by Lee et al.~\cite{lee2013crowdturfers}. 
	Their method relies on extracting features, which are used to train and test supervised machine learning (ML) classifiers. 
	The features extracted were associated with account properties, activity patterns, and linguistic properties.
	
	In 2015, another crowdturfing detection approach was developed by Song et al.~\cite{song2015crowdtarget}. 
	This approach identifies artificially promoted objects (such as posts, pages, and URLs), rather than crowdturfers or bots' accounts. 
	They described two types of malicious service provider websites: crowdturfing and black market websites. 
	Black market websites usually operate a large number of bots to perform promotion tasks in a given period of time; these bots resemble human accounts, however they operate as a group, and their activities are synchronized, performing the same task with a single deadline. 
	On the other hand, crowdturfing websites offer crowdturfers, which are either humans or advanced human-like bots, to execute their tasks. 
	Song et al. found that the accounts of crowdturfers are more popular than normal accounts, and their activities are not synchronized. 
	Later, they extracted features from retweeting patterns and used supervised ML algorithms to classify a tweet as artificially promoted or not.
	
	 Dickerson et al.~\cite{DBLP:conf/asunam/DickersonKS14} introduced an approach for the detection of bots which also used supervised ML classifiers based on features extracted from sentiment analysis, social network analysis, posted content, and account properties.
	
	In 2016, Davis et al.~\cite{Davis:2016:BSE:2872518.2889302} presented BotOrNot, a bot identification platform available through a Web user interface. 
	BotOrNot evaluates different aspects of Twitter accounts, including: 
	social network connections, 
	account properties, 
	content, and 
	behavioral features, as well as 
	sentiment analysis.
	
	Also in 2016, another bot detection method was introduced by Morstatter et al.~\cite{morstatter2016new}. 
	In contrast to other similar studies \cite{lee2013crowdturfers,DBLP:conf/asunam/DickersonKS14,song2015crowdtarget,Davis:2016:BSE:2872518.2889302}, their proposed method focused on a new supervised learning algorithm rather than feature extraction. 
	Their classifier was based on AdaBoost \cite{hu2008adaboost}, a known boosting algorithm, which is intended to improve the F1 measure, maintaining a balance between precision and recall.  
	Morstatter et al. used the latent Dirichlet allocation (LDA) topic probability distributions as features to test their proposed approach.

%
%
%
	
	The activities of accounts within OSM are naturally reflected in the content of their posts. 
	Some accounts publish posts mainly about cars, football, or movies. 
	Merchandise or service advertisements spread by company accounts also have their own characteristic vocabulary.  
	In contrast, accounts that ``retweet for hire'' may have no topical preference. 
	LDA~\cite{Blei:2003:LDA:944919.944937} can be used to identify the topics discussed in a large corpus of textual documents. 
	Morstatter et al.~\cite{morstatter2016new} treated each account as a document consisting of the concatenation of all of the content of its tweets.
	The LDA model provided a probability distribution for each account for each topic.
	These probabilities were used as features to train an account classifier.  
	
	Although topic detection was used in the past to assist in the identification of OSM \emph{abusers}, to the best of our knowledge, the current work is the first to focus on the detection of topics prone to OSM abuse. 

	While the prior studies mentioned above introduce novel approaches to detect OSM abusers and/or promoted posts, none of them provides a mechanism to measure the impact that these OSM abusers have on online discussions.
	
	Studies, such as~\cite{gayo2013predicting} and \cite{ferrara2016detection} focus on information credibility in OSM.
	Castillo et al.~\cite{gayo2013predicting} studied the propagation of false rumors on Twitter during a crisis event. 
	This approach is composed of three parts: emerging event detection, manual labeling, and feature extraction.
	
	The first part consists of collecting \emph{information cascades} which are collections of messages related to specific events.
	This step relies on a third party tool called Twitter Monitor,\footnote{http://www.twittermonitor.net/} which receives a query composed of keywords and logical propositions as input and returns the relevant set of messages.
	These cascades are then manually labeled using the crowdsourcing Mechanical Turk service.
	Finally, several factors observed in OSM are used to derive features, such as sentiments expressed in tweets, profile characteristics, and several other known features.
	
	Ferrara et al. \cite{ferrara2016detection}, proposed a ML framework to detect promoted campaigns and separate them from organic campaigns.
	In this study, topics are grouped by hashtags, which suggests that this method is applicable only to Twitter or to an OSM that uses similar tags.
	A number of features are presented, including network and diffusion-based, user account (profile) properties, sentiment features, and content-based features.

	While \cite{gayo2013predicting} and \cite{ferrara2016detection} tackle the problem of information credibility in OSM, we have identified the following limitations: both approaches use sentiment analysis which make them language dependent. Moreover, the need for information about the online profile (i.e., number of friends and followers) limits the applicability of these solutions to a specific OSM platform, since each platform possesses different properties for describing their profiles.
	On the other hand, in our approach four out of five of the proposed similarity functions rely on text properties, such as term probability distributions, term co-occurrences, etc., making our framework capable of working in multiple languages and on different OSM platforms.

	
	\section{Proposed Method} 
	\label{sec:genuineness_of_social_discussion}
	
\begin{figure}[htb]
		\centering
		\includegraphics[width=0.8\columnwidth]{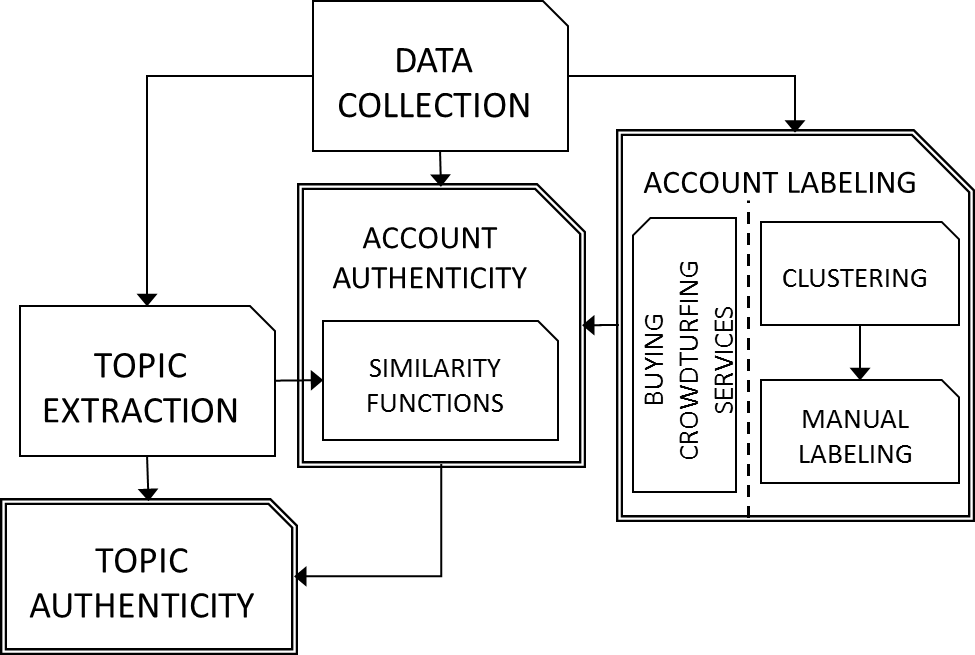}
		\caption{Estimation of topic authenticity}
		\label{fig:method}
	\end{figure}

	In this paper, we propose an approach for discriminating between online discussions artificially promoted by crowdturfers and those that attract the authentic interest of OSM accounts.
	In order to identify malicious manipulation of online discussions, we propose an approach based on estimating the prevalence of \emph{abusers} among the OSM accounts, which contributed to the given online discussion. 
	Our general approach is presented in Figure~\ref{fig:method}. 
	The steps performed, beginning with data collection and concluding with an estimation of topic authenticity, are described below.

The first step is the data collection process.
A discussion of the methods for collecting OSM data are out of the scope of this paper. 
The datasets used in this study are described in Section~\ref{sec:datasets}.  	

In the topic extraction stage for identifying online discussions we use a topic detection algorithm called LDA with hyperparameters optimized for the targeted OSM platform. 
A different algorithm like latent semantic analysis can also be used.
It is also possible to identify specific online discussion topics using predefined sets of keywords.

While executing the topic extraction, the account labeling process can execute simultaneously in order to obtain a ground truth labeled OSM account dataset.
For a more in depth description of the account labeling process, including a review of author types and the manual labeling guideline's definitions, see Section~\ref{sec:account_labeling}.
The labeled accounts and the topics are used later in order to quantify the authenticity of the OSM accounts based on several similarity functions (see Section~\ref{sec:author_similarity_functions}).
Later, we present the authenticity distribution for every topic in Section~\ref{sec:authenticity_of_accounts_and_topics}.


	\subsection{Account Labeling }
	\label{sec:account_labeling}
We propose two alternative approaches for creating labeled datasets of OSM accounts.
The first labeling approach is based on clustering the accounts, in order to have the highest variability among accounts, and then selecting a sample from each cluster; labels are assigned to the samples after manually inspecting them. 
We faced two primary challenges associated with this approach: the need for strict, unambiguous labeling guidelines and selection of accounts for labeling.
The second approach is based on acquiring a ground truth collection of OSM accounts that are known, with absolute certainty, to participate in crowdturfing platforms. 
In the following section, we present the guidelines for labeling accounts manually (see Section~\ref{sec:manually_labeling_process} and explain the manner for selecting accounts for labeling (see Section~\ref{clustering})).

	\subsubsection{Manual Labeling Guidelines} 
	\label{sec:manually_labeling_process}	
	
	\begin{figure*}
		\vspace{-0.5cm}	
		\centering
		\includegraphics[width=0.8\linewidth]{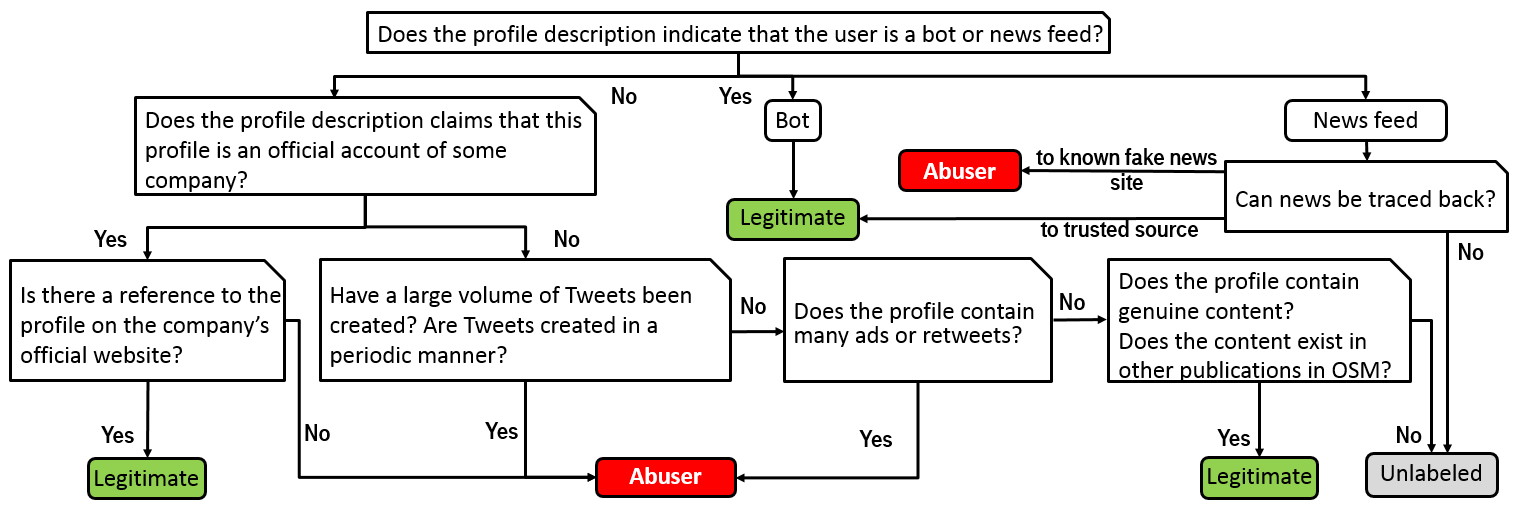}
		\caption{\vspace{-0.3cm}The manual labeling process.}
		\label{fig:labeling}
	\vspace{-0.6cm}	
	\end{figure*}

	We established the following guidelines for the manual classification of OSM accounts as \emph{abusers} or \emph{legitimate accounts}. 
	Figure~\ref{fig:labeling} presents the manual labeling process.
	Strict adherence to these clear labeling guidelines will ensure a high quality dataset.  
		
	(1) If the account explicitly mentions that it is a bot in its profile description, then we assume it is a benign automated account. 
	We mark it as \emph{legitimate}, because the account does not deceive OSM users regarding its nature. 

	(2) If an OSM account declares itself as a \emph{news feed} or any other type of content aggregator, we attempt to verify its authenticity. 
	If the news source is determined to be untrusted or is not a major news site, we label it as an \emph{abuser}, as opposed to a \emph{legitimate account}. 

	(3) If the account claims to be a company profile, this claim is validated as follows: 
	(3.1) Check to see that the account description links to an official website of the company. 
	(3.2) Look for a reference to the investigated Twitter account on the company's official home page, contact us page, press releases, or product pages. 
	If such a reference is found, the account is \emph{legitimate}. 
	Otherwise, it is considered a deceptive account and marked as an \emph{abuser}. 
	
	(4) Next, we inspect the account's behavior. 	
	If there are clear signs of automation, e.g., posting every five minutes twenty-four hours a day, or the account exhibits behavior not possible for a human user, we label it as an \emph{abuser}.

	(5) If there are many advertisements or retweets with marketing information alongside neutral or personal-looking posts, this is considered evidence that the account is involved in spam or crowdturfing activities. 
	It is important to note that some users install programs that post on their behalf; 
	in such cases, the account exhibits both human-like and non-human behavior patterns. 
	OSM accounts operated by naive users who grant permissions to untrusted third party applications are easy targets for spammers and crowdturfing platforms. 
	Thus, if the OSM account shows some human behavior, but also has frequent automated retweets or many advertisements, we treat it as an \emph{abuser}. 
	
	(6) Finally, if the majority of the posts published by the account contain authentic content (for example, if traces of the same personal content are found in other online social networks), we label it as \emph{legitimate}. 
	Having reliable labels for \emph{legitimate accounts} is important for training some supervised machine learning classifiers.  
	
	Unfortunately, it is not always possible to determine with absolute certainty whether or not an OSM account is \emph{legitimate}.   
	If none of the conditions apply or we have any doubt regarding a given account it remains \emph{unlabeled}. 

	We used the committee of experts approach in order to reach an agreement on the account labels. 
	Four experts (students) participated in the manual labeling process. 
	Three of the experts independently reviewed the same groups of unlabeled accounts and analyzed their Twitter profiles and posts.
	The experts then assigned a label of either \emph{legitimate account} or \emph{abuser} to each account based on the guidelines mentioned (see Figure~\ref{fig:labeling}).
	In cases of complete or almost complete agreement among the experts, the label was set according to majority vote. 
	In cases of reasonable doubt the account was not labeled.

	\subsubsection{Account Selection for Manual Labeling}
	\label{clustering}

	After defining the manually labeled guidelines, the arduous process of manual labeling should begin.
	However, the question is, which accounts should be selected for labeling?    
	A small sample of labeled instances randomly selected from a dataset may not accurately reflect the given population. 
	In order to have the greatest amount of variability among the labeled accounts, we first cluster the OSM accounts and select a sample of accounts from each cluster. 

	Clustering the accounts can be performed based on features extracted from the OSM account profile.
	Several studies found that profile information is useful for the identification of \emph{abusers}~\cite{lee2013crowdturfers,chu2010tweeting}. 
	We used features that were reported to perform well in the past~\cite{lee2013crowdturfers,DBLP:conf/asunam/DickersonKS14,Davis:2016:BSE:2872518.2889302}.
	These features include:
	the account age, number of followers, number of friends, friend-follower ratio, total number of posts, whether the profile configuration has been personalized or is the default setting, whether the profile image has been personalized or is the default setting, number of lists, whether the account has been verified by the social network, length of the screen name, average minutes between posts, average posts published during the total lifetime of the account, average posts published on days that the account has been active, and the number of retweets received.


	\subsubsection{Verified Abusers}
	\label{acquired_author_dataset}
	As an alternative to the manual account labeling, crowdturfing services can be used to obtain a sample set of verified \emph{abusers}.
	The idea is similar to the use of honeypots for attracting spammers, but in this case we actively invite \emph{abusers} to fall into the trap.
	
	First, we create several public Twitter accounts, which will serve as honeypots, and publish posts that contain clearly identified unique keywords. 
	Then, we buy followers and retweets from major crowdturfing sites, such as: \emph{intertwitter.com}, \emph{fastfollowerz.com}, \emph{socialshop.co}, \emph{socioblend.com}, \emph{coincrack.com}, and \emph{retweets.pro}.
	Since the accounts we created have no online activity, a short while after engaging the crowdturfing sites, all of the accounts' followers are crowdturfers as well. 
	We used Twitter API service in order to find users who cited the posts published by the honeypots and mark them as \emph{abusers}. 
	When taking this approach for collecting \emph{abuser's} data, there are a few points to keep in mind:
	(1) keywords contained in the posts published by honeypot accounts must be unique in order to avoid false labeling, 
	(2) the number of acquired followers and retweets should be modest enough to remain under the radar of OSM trend analysis engines and various content aggregators, and 
	(3) some of the accounts used for crowdturfing are short-lived and may be blocked by OSM providers 1-3 weeks after they are used for crowdturfing; 
	we have chosen to ignore such accounts in order to focus on the more advanced \emph{abusers} who try to avoid being detection.


	\subsection{Account Similarity Functions}
	\label{sec:author_similarity_functions}
	In order to quantify the authenticity of OSM accounts, we must introduce the following definitions. 
	Let \(A\) denote the set of accounts in the dataset. 
	Let \(P\) denote the collection of posts published by these accounts. 
	\(P(x)\subseteq P\) denotes the collection of posts published by account \(x\in A\).  
	For every post \(p\in P\), \(A(p)\) denotes its author. 
	We consider two posts equal (\(p_x = p_y\)) if their text is equal after stemming and stop words removal.	
	In order to estimate the authenticity of accounts we investigated multiple similarity functions.
	The similarity functions are defined over pairs of accounts \(f:A^2\rightarrow[0,1]\). 
	A good similarity function \(f(x,y)\) will return a low value if \(x\in A\) and \(y \in A\) belong to different classes, e.g., \(x\) is an \emph{abuser} and \(y\) is \emph{legitimate}, or vice versa.  
	Similarly, a high value of \(f(x,y)\) should indicate, with high probability, that \(x\) and \(y\) belong to the same class.
	In this study we used five different similarity functions: common-posts, topic-distribution, profile-properties, behavioral-properties, and bag-of-words.

	\begin{example}
	Consider, for example, the OSM accounts and posts presented in Figure~\ref{fig:input_info}. 
	Accounts (ovals) point to their posts. 
	Accounts may have properties associated with them, e.g., the number of friends and followers, and the time since the account was created.  
	A2 was labeled as an \emph{abuser}, and A5 is a \emph{legitimate} account. 
	A1, A3, and A4 are unlabeled. 
	We use this example to demonstrate the  similarity functions. 
	\begin{figure}[htb]
		\vspace{-0.3cm}	
		\centering
		\includegraphics[width=1\linewidth]{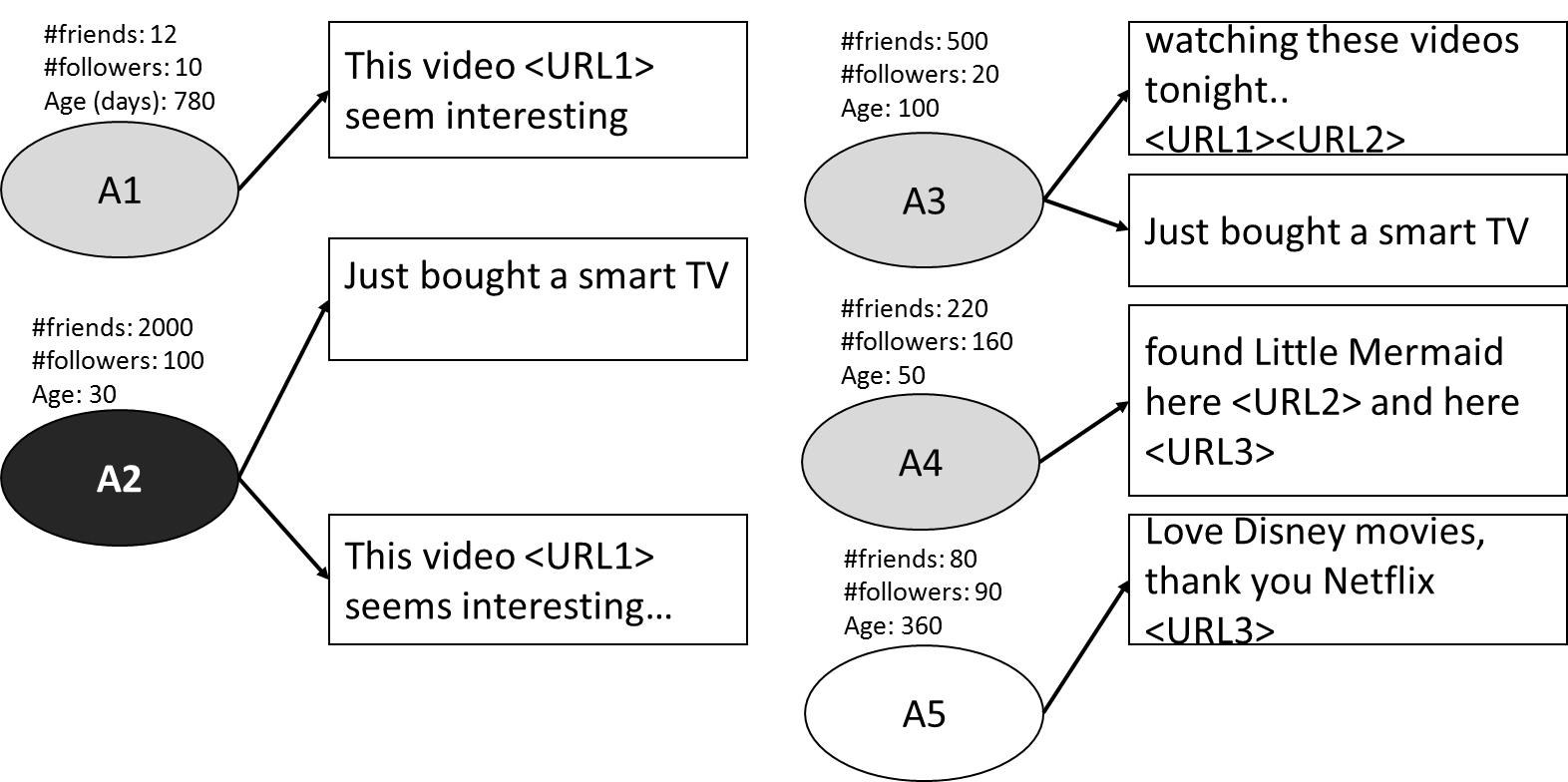}\vspace{-0.3cm}
		\caption{Illustration for understanding the similarity functions}
		\label{fig:input_info}
	\vspace{-0.2cm}	
	\end{figure}
	\end{example}

%
%
%
%
%

	\subsubsection{Common-Posts}
	OSM accounts that publish the same content might be part of a crowdturfing campaign~\cite{lee2015characterizing,zhang2012detecting}. 
	This similarity function shows which accounts spread the same content across the OSM.
	We use the Jaccard coefficient to normalize the number of common posts.
	\[ common\text{-}posts(x,y) = \frac{ \left| P(x)\cap P(y) \right| } { \left| P(x)\cup P(y) \right| }
	\]
	\begin{example}
	Consider the example in Figure~\ref{fig:input_info}. A2 has one post in common with A1 (despite the minor differences in the posts' text).  
	\end{example}
	
	\subsubsection{Bag-of-Words}
	Common posts may work well for identifying groups of cooperating spammers or simple bots used to promote a product or website through search engine optimization. 
	However, this function may fail to detect human operated accounts that are employed in the same campaign due to text variability in the crowdturfers' posts. 
	Nevertheless, \emph{abusers} that participate in the same crowdturfing campaign would likely use roughly similar vocabularies.

	In order to quantify the similarity of vocabularies used by different accounts we employ the \emph{bag-of-words} approach.
	According to this approach, all the content posted by a given account is transformed into a set of terms.
	We denote the set of stemmed terms used in a post \(p\in P\) as \(W(p)\).
	We assume that \(W(p)\) does not contain stop-words.
	We define the vocabulary \(W(x)\) of an account \(x\in A\) as the collection of all terms used by this account in the given dataset \(W(x)=\bigcup_{p\in P} W(p)\). 
	We use Jaccard coefficient to quantify the similarity between the vocabularies used by two accounts: 	
	\[
	 bag\text{-}of\text{-}words(x,y) = \frac{|W(x) \cap W(y)|}{|W(x)\cup W(y)|}
	\]
	Hashtags and URLs are considered as terms for this purpose and are not stemmed.

	\begin{example}
	Consider accounts A1 and A3 in Figure~\ref{fig:input_info}. 
	The term \emph{video} and URL1 are used by both. 
	\emph{This, these, a, just, seem} are stop words, leaving nine terms, including URLs, which were used by either A1 or A3.  
	Thus, the bag-of-words similarity between them is: 
		$bag\text{-}of\text{-}words(A1,A3) = \frac{2}{9} $ 
	\end{example}	

	\subsubsection{Topic-Distribution}
	The next similarity function compares the topical distributions of posts published by each account. 
	While the \emph{bag-of-words} roughly compares which terms are used by two accounts, here we consider the usage frequencies as well.  
	
	
The LDA algorithm for topic detection determines the probability of each vocabulary term to be used in each topic. 
Based on these term-in-topic probabilities LDA-based topic detection determines the probability \(T_{p,i}\) that the post \(p\) belongs to topic \(i\). 
Next, for each OSM account $x$ and each topic \(i\), we compute the average probability that \(x\)'s posts belong to topic \(i\). 
\[T'_{x,i}=avg_{p\in P(x)} \left\{ T_{p,i} \right\}\vspace{-0.0cm}
\]
	
	Finally, for each pair of accounts $x, y$, we measure the similarity between  vectors \(T'_x\) and \(T'_y\) using cosine similarity. \
	\[ topic\text{-}distr(x,y) = cosine\_similarity(T'_x,T'_y)\]
	
	Although any kernel function can be used to compare vector representations of accounts, we only use cosine similarity in the rest of this section; 
	other kernel functions we tried during this study produced similar results. 

	\subsubsection{Profile Properties}
	Accounts can be compared based on the features extracted from their OSM profiles. 
	We use profile features that were shown to help in account classification in the past~\cite{lee2013crowdturfers,DBLP:conf/asunam/DickersonKS14,Davis:2016:BSE:2872518.2889302}.
	The features are listed in Section~\ref{clustering}. 
	Let \(\overrightarrow{x_p}\) and \(\overrightarrow{x_p}\) represent the profile feature vectors of accounts \(x\) and \(y\), respectively. 
	We use cosine similarity to compare these two vectors. 
\[ profile\text{-}prop(x,y) = cosine\_similarity(\overrightarrow{x_p},\overrightarrow{y_p})\]
	
	\begin{example}
	Examples of profile features are presented above the account names in Figure~\ref{fig:input_info}.
	For example, A1 has 12 friends and 10 followers, and it is 780 days old. 
	According to these features, A1 is more similar to A5 than to A2, having \(profile\text{-}prop(A1,A5) = 0.95\) and \(profile\text{-}prop(A1,A2) = 0.03\), respectively.
	\end{example}
	
	\subsubsection{Behavioral-Properties}
	Techniques relying on features that describe account behavior were also useful for the identification of OSM \emph{abusers}~\cite{lee2013crowdturfers,DBLP:conf/asunam/DickersonKS14}. 
	These features include:
	total number of retweets, average number of retweets, average number of hashtags, average number of hyperlinks, average user mentions, and average post length. 
	Similar to \(profile\text{-}prop\), we use \(cosine\_similarity\) to compare the account behavior feature vectors. 
	\[behavior\text{-}prop(x,y) = cosine\_similarity(\overrightarrow{x_b},\overrightarrow{y_b}) 
	\]
	where $\overrightarrow{x_b}$ and $\overrightarrow{y_b}$ are vectors of features describing the account behavior.


\subsection{Authenticity of Accounts and Topics}
\label{sec:authenticity_of_accounts_and_topics}
	
In order to quantify the authenticity of OSM accounts in the dataset, we compare them to labeled accounts (discussed in Section~\ref{sec:account_labeling}) using the similarity functions presented in Section~\ref{sec:author_similarity_functions}. 
The simplest and most intuitive algorithm that can be applied in these settings is k-nearest neighbors (KNN). 
According to this algorithm, for each unlabeled account \(x\), we select the \(k\) most similar labeled accounts (nearest neighbors). 
Then, \(x\) receives the label of the majority of the nearest neighbors. 
This approach results in quite accurate classification as will be discussed in Section~\ref{sec:evaluation}. 
Nevertheless, for the purpose of this study, binary classification is not always sufficient. 
Since many crowdturfers are labeled as \emph{abusers}, but they are actually behave partially legitimate, we can use the classification confidence as the measure for the authenticity of an account. 

More specifically, we define the authenticity of the account \(x\) as the confidence in \(x\) being a \emph{legitimate} account. 
Since this paper focuses on account similarity functions, we choose various KNN confidence measures~\cite{dalitz2009reject}. 
The choice of the specific confidence measure is guided by the characteristics of our problem domain.  
First, the number of labeled accounts is rather limited, especially if we choose the manual labeling approach. 
Second, the similarity to the labeled accounts contains important information that should be considered. 
Finally, we acknowledge the fact that there are many sub-types of \emph{abusers} and \emph{legitimate accounts}. 
This is clearly indicated by the multi-step manual labeling process presented earlier. 
Thus, we should not rely on the similarity of the tested account to distant labeled instances of each class, but only consider the similarity to the nearest neighbors.     

We choose the confidence measure similar to the one defined by Arlandis et al.~\cite{arlandis2002rejection},
  
\[acc\text{-}auth(x) =0.5 -  \dfrac{\sum_{y \in N^k_{abuser}(x)} f(x,y)  }{  \sum_{y\in N^k(x) } f(x,y) }\]
Where \(N^k(x)\) is the set of \(x\)'s nearest neighbors, \(N^k_{abuser}(x)\subseteq N^k(x)\) is the set of known abusers among the nearest neighbors, and \(f\) is the account similarity function. 
The constant factor is set to 0.5 in this equation in order to allow negative, as well as positive authenticity values (\(authenticity(x)\in [-0.5,0.5]\)).  

The last step of the proposed approach is aggregating the authenticity of individual OSM accounts into authenticity of topics.  
While there are multiple aggregation options, we consider the following two aggregations: 
\paragraph{Post level aggregation} 
In order to form the topic authenticity, every post is associated with the authenticity of its author. 
The authenticities of the posts which related to a specific topic are accumulated in the topic authenticity. 

\begin{equation}
\label{eq:post-distr}
topic\text{-}auth\text{-}1(i)=\sum_{p\in P} T_{p,i}\cdot acc\text{-}auth(A(p)) 
\vspace{-0.2cm}
\end{equation}

\paragraph{Author level aggregation} 
First, a set of authors of posts for a specific topic is determined.
Then, authenticities of the author accounts are aggregated. 
We define the set of accounts involved in a specific topic \(i\) as 
\vspace{-0.2cm}
\[D(i)=\left\{ A(p) : T_{p,i} = MAX_j\left\{T_{p,j}\right\}  \right\}.\vspace{-0.2cm}\] 
Here, every post is associated with a single topic -- the one it belongs to with the highest probability. 
An account is associated with a topic if at least one of its posts is associated with that topic. 
The account level authenticity of the topic \(i\) is then determined as follows:  
\begin{equation}
\label{eq:auth-distr}
topic\text{-}auth\text{-}2(i)=\sum_{x\in D(i)} acc\text{-}auth(x)
\vspace{-0.2cm}
\end{equation}

We do not distinguish between authentic and malicious posts published by the same account. 
Thus, in the case of partially automated accounts and casual crowdturfers, even topics in which they express authentic interest are tainted by their illegitimate activities. 
Future work may consider specific post classification into legitimate activities or activities that are part of a crowdturfing campaign. 
We argue that information contained in a single post is often insufficient for performing accurate classification, especially in the case of human operated crowdturfing accounts.  

The proposed post level topic authenticity has higher granularity and is more robust, because it considers the affinity of a post to every topic. 
In contrast, the author level aggregation may introduce more noise in small datasets, because it disregards the lesser post to topic affinities.
It means that the author level topic authenticity is affected only by a single topic as opposed to post topic level authenticity which is measured by each topic. 
Nevertheless, author level aggregation is important, because it reduces the weight of accounts aggressively publishing on a specific topic.


	\section{Evaluation} 

	\label{sec:evaluation}
	In this section we evaluate the investigated account similarity functions on three Twitter datasets and demonstrate the topic authenticity estimation.

	\subsection{Datasets}
	\label{sec:datasets}
	
	Three datasets were used for evaluation of the author similarity functions: \emph{Manually Labeled Accounts}, \emph{Verified Abuser}, and \emph{Arabic Honeypot datasets}.
	The first two datasets are associated with \emph{virtual TV domain}. 
	These two datasets contain the same unlabeled accounts and their posts.
	However, their labeled accounts are different.   
	The \emph{Arabic Honeypot dataset}~\cite{morstatter2016new} was used for validation of the results.

	\subsubsection{Manually Labeled Accounts Dataset}
	\label{sec:manually_labeled_accounts_dataset}
	
	For this study we collected Twitter data over a period of five months (from April 25 to September 19, 2016), crawled with the help of VICO Research \& Consulting GmbH~\cite{vico}.
	We collected tweets containing one of the following key phrases: `\emph{Online TV},' `\emph{Internet TV},' and `\emph{Smart TV}.'
	This domain was chosen because it contains many tweets published by both \emph{abusers} and \emph{legitimate accounts}. 
	
	We used the Twitter REST API~\cite{twitter_rest_api} to collect the profile features, such as number of statuses, number of followers, number of friends, etc.
	The \emph{Virtual TV Manually Labeled Account dataset} contain 89,111 accounts, and 188,211 tweets that were published by these accounts.
	Following the manual labeling process presented in Section~\ref{sec:manual_labeling_guidelines}, we identified 289 \emph{abusers} and 316 \emph{legitimate accounts}. 
	
	\subsubsection{Verified Abuser Dataset}	
	\label{sec:verified_abusers_dataset}
	
	In order to form the set of \emph{verified abusers}, we created nine public Twitter accounts that published a total of 53 posts.
	Using crowdturfing sites, we hired 3,169 Twitter \emph{verified abusers} to follow those accounts and/or retweet their posts.
	These accounts are defined as \emph{verified abusers}, because they make receive financial incentives to perform activities like following accounts and retweeting posts.
	1,968 of the \emph{verified abusers} were blocked by Twitter during the period of our study; 
	therefore, these accounts were not included in the current study because, with the exception of the posts that we paid for, the data does not include their posts or profile information. 
	The other 1,201 accounts were still active at the time of this paper's writing. 
	
	Four out of five similarity functions used in this study rely on posts. 
	Thus, we disregard 195 accounts, which published less than thirty posts in their lifetime.
	Overall, our dataset contains 1,006 active Twitter accounts of \emph{verified abusers} with more than thirty posts each.
	In addition, the dataset also contains 1,068 additional accounts that posted about the \emph{virtual TV} domain. 
	We consider these accounts as \emph{legitimate} for the sake of this study.

	\subsubsection{Arabic Honeypot Dataset}
	\label{sec:arabic_honeypot_dataset}
	
	In addition to the Twitter datasets described above we used the \emph{Arabic Honeypot dataset} provided by Morstatter et al.~\cite{morstatter2016new}. 
	This data was collected using a honeypot network.
	Morstatter et al. reported that the dataset contains 6,285 accounts, and 725,179 Twitter posts published by those accounts.
	Because this dataset only contains the Twitter account ID and Twitter post ID, the Twitter REST API service was used in order to obtain the missing information about the accounts and posts.
	Not that only 5,270 Twitter accounts out of the 6,285 are still operational when this study was conducted.
	Next, we filtered out 1,144 of the 5,270 accounts, because either the accounts were suspended, the accounts had less than thirty posts for analysis, or the privacy settings were changed to protected, preventing us from collecting their private posts.  
	Overall, we collected 652,128 posts from 4,126 accounts, among them 2,042 \emph{abusers} and 2,084 \emph{legitimate accounts}.
	A summary of all of the datasets used in this study is presented in Table \ref{tab:datasets}.
	
	\begin{table*}[!ht]
		\caption{Datasets Summary}
		\begin{center}
			\begin{tabular}{cccccc}
				Domain & \specialcell{Collected \\ Accounts} & \specialcell{Collected \\ Posts} & \specialcell{Labeling \\ Method} & \specialcell{ Legitimate Accounts} & \specialcell{Abusers} \\
				\hline
				\multirow{2}{*}{Virtual TV}&\multirow{2}{*}{89,111}&\multirow{2}{*}{188,211}& Manual labeling & 316 & 289 \\\cline{4-6}
				& & & Buying crowdturfing services & 1,068 & 1,006\\
				\hline
				Arabic Honeypot & 4,126 & 652,128 & honeypot  & 2,084 & 2,042 \\
				\hline
			\end{tabular}
		\end{center}
		\label{tab:datasets}
	\end{table*}

	\subsection{Account Classification}
	\label{sec:account_classification}
	We evaluate the account similarity functions presented in Section~\ref{sec:author_similarity_functions} by applying a KNN classifier (with ($1\leq k \leq 5$).
	The size of the training set varied from 1 to 40\% of the labeled accounts in each dataset. 
	Every evaluation round was executed five times with different randomly selected training sets.  
	The performance indicators used are the Area under ROC curve (AUC) and F1 measures.

\begin{figure}[!htb]
		\centering
		\includegraphics[width=1\linewidth, height=1.5in]{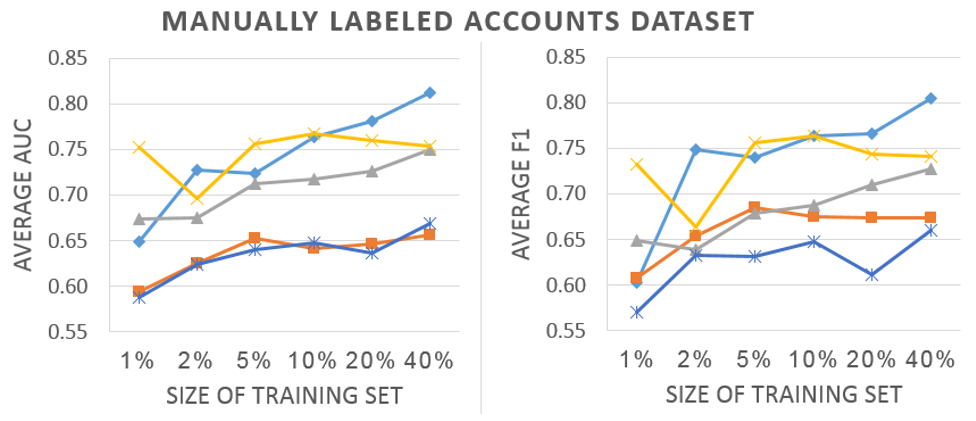}
		\label{fig:result_manual_label}
		\centering
		\includegraphics[width=1\linewidth, height=1.5in]{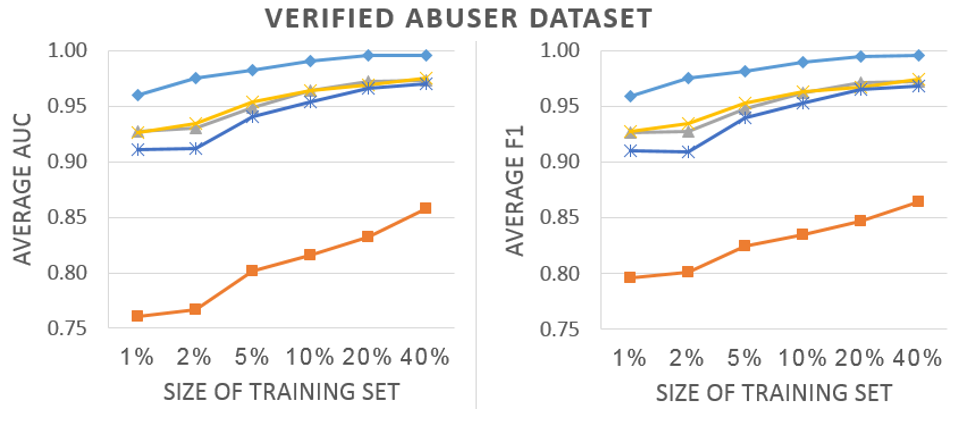}
		\label{fig:results_acquired}
		\centering
		\includegraphics[width=1\linewidth, height=1.5in]{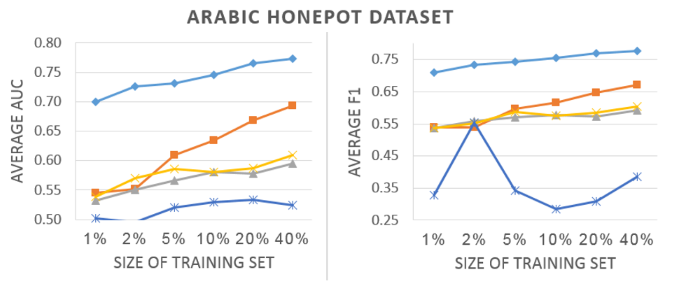}
		\label{fig:results_asu_honeypot}
		\centering
		\includegraphics[width=1\linewidth]{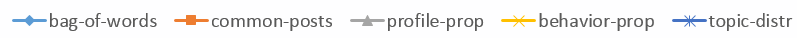}
		\caption{ Average AUC and F1 scores obtained with the three datasets}
		\label{fig:legend}

	\end{figure}

	Figure~\ref{fig:legend} presents the average performance of the similarity functions for the \emph{Manually Labeled Accounts}, \emph{Verified Abuser}, and \emph{Arabic Honeypot datasets}.
	The x-axis corresponds to the percent of labeled accounts used as a training set. 
	The y-axis corresponds to the AUC (left) and the F1 measure (right).

	In all of the datasets, the bag-of-words was found to be the superior similarity function according to both AUC and F1 measures. 
	It also has the lowest standard deviation for both AUC and F1 ranging from 0.002 to 0.016 across all datasets, and 
	it outperforms both the simpler common-posts similarity function and the more sophisticated topic-distribution similarity function. 
	It is interesting to note that common-posts is the worst performing similarity function for detecting the \emph{abusers} that are employed with the crowdturfing platforms. 
	This observation strengths the idea that accounts used for crowdturfing (some of which are human operated) cannot be detected easily by publishing the same content.    
	However, these accounts are detected with a high degree of accuracy by all of the other similarity functions, with the bag-of-words reaching AUC, and F1 measures of 0.996 with about 800 accounts in the training set (40\%).
	Profile properties (e.g., number of followers, account age) and behavioral properties (e.g., number of hashtags) perform reasonably well, achieving similar results in all datasets. 
 	The worse similarity function in the Arabic Honeypot dataset was \emph{topic-repr}. 
	While surprised by this result, we attribute to the poor performance of standard LDA based topic detection algorithms on the Arabic language. 
	Surprisingly, high performance of the \emph{common-posts} similarity function suggests that the number of automated accounts is relatively high in this dataset.  
		
	\subsection{Topic Authenticity} 
	\label{sec:measuring_genuineness_of_social_discussion}
	
	We measured the topic authenticity for the \emph{virtual TV dataset} described in Section~\ref{sec:datasets}.
	The Arabic Honeypots data set was not included in this analysis because (1) the researchers lack knowledge of the Arabic language and could not qualitatively validate the calculated topic authenticity, and (2) LDA seemed to perform poorly on this dataset without proper tuning. 
	The number of topics in the collected Virtual TV data was optimized empirically to produce 48 coherent topics.  
	Based on the \emph{Manually Labeled Accounts} training set, we computed an authenticity score for each unlabeled author in the dataset. 
	The bag-of-words similarity function was used for this purpose, since it provided the best performance in terms of the AUC and F1 measures.

\begin{figure}[!htb]
		\centering
		\includegraphics[width=1\linewidth]{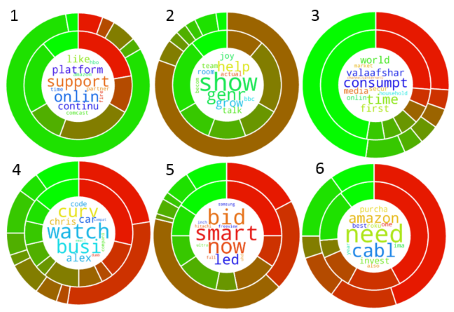}
		\caption{Authenticity distribution in six topics from the Virtual TV dataset}
		\label{fig:topic_visualization}
	\end{figure}

	In order to visually represent the authenticity of each topic we used donut charts as depicted in Figure~\ref{fig:topic_visualization}.
	For the sake of brevity, we show the authenticity distribution of just six of the forty-eight topics found in the \emph{Manually Labeled Accounts dataset}. 		
	In the middle of each donut chart we include the word cloud representing the topic. 
	Words are sized based on their probability of being associated with the given topic. 
	The inner circle of the donut chart enclosing each word cloud represents the account authenticity distribution as presented in Equation~\ref{eq:auth-distr}.
	Similarly, the outer circle depicts the post level authenticity distribution as presented in Equation~\ref{eq:post-distr}.  

	It can seen that in some cases (e.g., topics 1,2, and 4) the fraction of posts is disproportional to the fraction of accounts having the same authenticity level. 
	This means that a few accounts (\emph{legitimate accounts} in the case of topic 1 and \emph{abusers} in the case of topic 4) took over the discussion in these topics and may have had a strong influence on the other accounts who wrote on this topic.   
	Overall, it is easy to distinguish between topics in which most of the participating accounts are \emph{legitimate} and topics overwhelmed with \emph{abusers}. 
	For example, the online discussions about online support platforms seem to be authentic. 
	In contrast, topics mentioning Amazon and bidding are highly promoted by OSM abusers. 
	Statistical analysis of the authenticity distributions using ANOVA identified groups of online discussions that contain commercial content (largely prone to \emph{abusers}) and groups of online discussions that raise little financial interest (mostly authentic). 
	These results show that the proposed analytical pipeline for topic authenticity estimation makes sense, confirming that the suggested approach to detect OSM manipulation in politics and various other domains can be used in future studies.



	\section{Ethical Considerations} \label{ethical}
	Collecting data from OSM has raised ethical concerns in recent years~\cite{elyashar2013homing,Boshmaf:2011:SNB:2076732.2076746,elyshar2012organizational,elyashar2014guided}.
	In order to minimize the potential risks that may arise from such activities, this study follows recommendations presented by \cite{flicker2004ethical} and \cite{elovici2014ethical}, which deal with ethical challenges regarding OSM and Internet communities.
	
	In this study VICO Research \& Consulting GmbH provided publicly available posts from Twitter. 
	To obtain missing information about accounts and their posts, we used the Twitter REST API which only provides public information.
	This means that we do not collect the information of accounts that do not agree to share their information publicly.
	
	The Arabic Honeypot dataset made available to the research community lacks the identifiable properties of accounts; in addition, in the published dataset we replaced all tweets that appear less than 20 times on Twitter with words randomly selected from the topic's word distribution.  

	The research protocol was approved by the Ben-Gurion University of the Negev Human Research Ethics Committee. 
	
	\section{Conclusions and Future Work} 
	\label{conclusions}

	In this study, we demonstrated topic authenticity on three Twitter datasets.
	Based on the results we can draw the following conclusions: First, the similarity function that provided the best performance was the \emph{bag-of-words}, which obtained the highest average AUC and F1 and the lowest standard deviation for each of the datasets. 
	These results demonstrate that the \emph{bag-of-words} is the optimal similarity function for measuring the accounts' authenticity among the five similarity functions we measured.

	Second, with regard to the AUC and F1 performance measures, 
	in the \emph{Verified Abuser dataset} we obtained to almost the maximal result, as opposed to the two other datasets which obtained AUC and F1 measures of approximately 0.8.   
	We believe that the high performance of the \emph{Verified Abuser dataset} was due to the composition of accounts provided by crowdturfing sites which were actually simple bots. 
	The characteristics of simple bots are likely easier to identify than the more sophisticated \emph{abusers} found in other datasets used in our evaluation. 
	
	Third, by using our proposed method we succeeded at differentiating between topics that are prone to OSM manipulation and topics that attract authentic public interest.
	Moreover, due to the fact that our proposed method includes a manual labeling process that contains clustering analysis, even with a small number of labeled \emph{abusers} we can detect manipulated topics. 

	
	
	In the future, we plan to evaluate the presented approach on additional datasets spanning multiple social networks, and compare the topic authenticity across multiple domains, such as politics, sports, etc.
	In addition, we think that it will be interesting to understand whether the \emph{abusers} follow the same pattern in other domains. 
	Moreover, we aim to incorporate the capability of detecting camouflaged abusers into our approach.
	We also would like to measure the use of the \emph{bag-of-words} for clustering analysis in order to improve the clusters provided .    

	\section*{Acknowledgment}
	
	The authors would like to thank Robin Levy-Stevenson for proofreading this article. 

	\bibliographystyle{IEEEtran}
	\bibliography{references}
	
\end{document}